# Negative kinetic energy term of general relativity and its removing

T. Mei

(Department of Journal, Central China Normal University, Wuhan, Hubei PRO, People's Republic of China
E-Mail: meitao@mail.ccnu.edu.cn meitaowh@public.wh.hb.cn )

**Abstract:** We first present a new Lagrangian of general relativity, which can be divided into kinetic energy term and potential energy term. Taking advantage of vierbein formalism, we reduce the kinetic energy term to a sum of five positive terms and one negative term. Some gauge conditions removing the negative kinetic energy term are discussed. Finally, we present a Lagrangian that only include positive kinetic energy terms. To remove the negative kinetic energy term leads to a new field equation of general relativity in which there are at least five equations of constraint and at most five dynamical equations, this characteristic is different from the normal Einstein field equation in which there are four equations of constraint and six dynamical equations.

In Ref. [1], it has been pointed out that there is a negative kinetic energy term in Lagrangian of the Einstein-Hilbert action of general relativity; however, this conclusion was obtained under the Schwinger "time gauge" $e^{0}_{\hat{a}}=0 \ (a=1,2,3)$ for tetrad $e^{\hat{\alpha}}_{\mu}$. In this paper, we discuss general case. All symbols and conventions follow Ref. [1].

## 1 Lagrangian of general relativity

As is well known, the Einstein-Hilbert action of general relativity is

$$S_{\mathrm{EH}}=\frac{c^3}{16\pi G}\int\sqrt{-g}\mathrm{d}^4xR=\frac{c^3}{16\pi G}\int\sqrt{-g}\mathrm{d}^4x\,L_{\mathrm{g}}+\frac{c^3}{16\pi G}\int\mathrm{d}^4x\frac{\partial}{\partial x^\mu}\left[\sqrt{-g}\,(g^{\rho\sigma}\Gamma^{\mu}_{\rho\sigma}-g^{\mu\rho}\Gamma^{\sigma}_{\rho\sigma})\right], \quad (1\text{-}1)$$

$$L_{\mathrm{g}}=g^{\alpha\beta}(\Gamma^{\rho}_{\alpha\sigma}\Gamma^{\sigma}_{\beta\rho}-\Gamma^{\rho}_{\alpha\beta}\Gamma^{\sigma}_{\rho\sigma}), \quad (1\text{-}2)$$

Especially, the quadratic term of time derivative in $L_{\mathrm{g}}$ is

$$\begin{aligned}L_{\mathrm{g}0}&=\frac{1}{4}\left[-2g^{0i}\left(g^{0l}g^{jm}-g^{0j}g^{lm}\right)+g^{00}\left(g^{ij}g^{lm}-g^{il}g^{jm}\right)\right]g_{ij,0}g_{lm,0}\\&=\frac{1}{4}\left(-g^{00}\right)\left(\tilde{g}^{il}\tilde{g}^{jm}-\tilde{g}^{ij}\tilde{g}^{lm}\right)g_{ij,0}g_{lm,0}\,.\end{aligned} \quad (1\text{-}3)$$

where $\tilde{g}^{ij}=g^{ij}-\frac{g^{0i}g^{0j}}{g^{00}}$, and $\tilde{g}^{ik}g_{kj}=\delta^i_j$.

Taking advantage of the inverse matrix $\tilde{g}^{ij}$ of $g_{ij}$, we can introduce 3-dimentional Christoffel symbol

$$\tilde{\Gamma}^i_{lm}=\frac{1}{2}\tilde{g}^{ij}\left(g_{jl,m}+g_{jm,l}-g_{lm,j}\right), \tag{1-4}$$

and we can prove easily that the relation between $\tilde{\Gamma}^i_{lm}$ and the 4-dimentional Christoffel symbol $\Gamma^\mu_{\rho\sigma}=\frac{1}{2}g^{\mu\nu}\left(g_{\nu\rho,\sigma}+g_{\nu\sigma,\rho}-g_{\rho\sigma,\nu}\right)$ is:

$$\tilde{\Gamma}^i_{lm}=\Gamma^i_{lm}-\frac{g^{0i}}{g^{00}}\Gamma^0_{lm}. \tag{1-5}$$

Based on $\tilde{g}^{ij}$ and $\tilde{\Gamma}^i_{lm}$, we define

$$L_{\mathrm{GK}}=\frac{1}{-g^{00}}\left(\tilde{g}^{il}\tilde{g}^{jm}-\tilde{g}^{ij}\tilde{g}^{lm}\right)\Gamma^0_{ij}\Gamma^0_{lm}, \tag{1-6}$$

$$\begin{aligned}L_{\mathrm{GU}}&=\tilde{g}^{ij}\left(\frac{g^{0\lambda}}{g^{00}}\left(\Gamma^0_{\lambda k}\tilde{\Gamma}^k_{ij}-\Gamma^0_{\lambda j}\tilde{\Gamma}^k_{ik}\right)-\left(\tilde{\Gamma}^l_{im}\tilde{\Gamma}^m_{jl}-\tilde{\Gamma}^l_{ij}\tilde{\Gamma}^m_{lm}\right)\right)\\&=\frac{g^{0\lambda}}{g^{00}}\Gamma^0_{\lambda k}\left(\tilde{g}^{ij}\tilde{\Gamma}^k_{ij}-\tilde{g}^{ik}\tilde{\Gamma}^j_{ij}\right)-\tilde{g}^{ij}\left(\tilde{\Gamma}^l_{im}\tilde{\Gamma}^m_{jl}-\tilde{\Gamma}^l_{ij}\tilde{\Gamma}^m_{lm}\right)\\&=\frac{1}{2g^{00}}g^{0\rho}g^{0\sigma}g_{\rho\sigma,k}\left(\tilde{g}^{ij}\tilde{\Gamma}^k_{ij}-\tilde{g}^{ik}\tilde{\Gamma}^j_{ij}\right)-\tilde{g}^{ij}\left(\tilde{\Gamma}^l_{im}\tilde{\Gamma}^m_{jl}-\tilde{\Gamma}^l_{ij}\tilde{\Gamma}^m_{lm}\right);\end{aligned} \tag{1-7}$$

We can prove that the remainder of $\sqrt{-g}L_{\mathrm{g}}$ and $\sqrt{-g}\left(L_{\mathrm{GK}}-L_{\mathrm{GU}}\right)$ is a total derivative:

$$\sqrt{-g}L_{\mathrm{g}}=\sqrt{-g}\left(L_{\mathrm{GK}}-L_{\mathrm{GU}}\right)+\frac{\partial}{\partial x^\mu}\left[\frac{\sqrt{-g}}{g^{00}}\left(g^{0\nu}\frac{\partial g^{0\mu}}{\partial x^\nu}-g^{0\mu}\frac{\partial g^{0\nu}}{\partial x^\nu}\right)\right]. \tag{1-8}$$

We therefore can employ

$$S_{(1)}=\frac{c^3}{16\pi G}\int\sqrt{-g}\mathrm{d}^4x\left(L_{\mathrm{GK}}-L_{\mathrm{GU}}\right) \tag{1-9}$$

as an action of general relativity. From (1-7) we see that there is not any time derivative term in $L_{\mathrm{GU}}$, hence, $L_{\mathrm{GK}}$ and $L_{\mathrm{GU}}$ can be regarded as kinetic energy term and potential energy term, respectively. Especially, the quadratic term of time derivative in $L_{\mathrm{GK}}$ is just (1-3).

The form (1-6) of $L_{\mathrm{GK}}$ has been given under the "time gauge" $e^0_{\hat{a}}=0$ $(a=1,2,3)$ for tetrad $e^{\hat{\alpha}}_\mu$ in Ref. [1], what we discuss here is general case.

Substituting $g_{\mu\nu}=e^{\hat{\alpha}}_\mu e_{\hat{\alpha}\nu}$ and $g^{\mu\nu}=e^\mu_{\hat{\alpha}}e^{\hat{\alpha}\nu}$ to (1-6) and (1-7), we obtain the vierbein form of $L_{\mathrm{GK}}$ and $L_{\mathrm{GU}}$ immediately; however, for below discussion, we rewrite $L_{\mathrm{GK}}$ to another form.

Defining $\tilde{e}^i_{\hat{a}} = e^i_{\hat{a}} - \frac{e^0_{\hat{a}} e^i_{\hat{0}}}{e^0_{\hat{0}}}$, which satisfies $\tilde{e}^i_{\hat{a}} e^{\hat{a}}_j = \delta^i_j$, $\tilde{e}^i_{\hat{a}} e^{\hat{b}}_i = \delta^{\hat{b}}_{\hat{a}}$. On the other hand, we have $\tilde{g}^{ij} e^{\hat{a}}_i e^{\hat{b}}_j = \eta^{\hat{a}\hat{b}} - \frac{e^{\hat{a}0} e^{\hat{b}0}}{g^{00}}$. Using these formulas, we obtain

$$L_{\mathrm{GK}} = \frac{1}{-g^{00}} \left( \tilde{g}^{pu} \tilde{g}^{qv} - \tilde{g}^{pq} \tilde{g}^{uv} \right) e^{\hat{a}}_p e^{\hat{b}}_q e^{\hat{c}}_u e^{\hat{d}}_v \left( \tilde{e}^i_{\hat{a}} \tilde{e}^j_{\hat{b}} \Gamma^0_{ij} \tilde{e}^l_{\hat{c}} \tilde{e}^m_{\hat{d}} \Gamma^0_{lm} \right) = \frac{1}{-g^{00}} Q^{\hat{a}\hat{b}\hat{c}\hat{d}} Y_{\hat{a}\hat{b}} Y_{\hat{c}\hat{d}}, \tag{1-10}$$

$$Q^{\hat{a}\hat{b}\hat{c}\hat{d}} = \left( \eta^{\hat{a}\hat{c}} - \frac{e^{\hat{a}0} e^{\hat{c}0}}{g^{00}} \right) \left( \eta^{\hat{b}\hat{d}} - \frac{e^{\hat{b}0} e^{\hat{d}0}}{g^{00}} \right) - \left( \eta^{\hat{a}\hat{b}} - \frac{e^{\hat{a}0} e^{\hat{b}0}}{g^{00}} \right) \left( \eta^{\hat{c}\hat{d}} - \frac{e^{\hat{c}0} e^{\hat{d}0}}{g^{00}} \right), \tag{1-11}$$

$$Y_{\hat{a}\hat{b}} = \tilde{e}^i_{\hat{a}} \tilde{e}^j_{\hat{b}} \Gamma^0_{ij}. \tag{1-12}$$

We can prove that the changes of the action (1-9) in which both $L_{\mathrm{GK}}$ and $L_{\mathrm{GU}}$ take vierbein form under coordinate transformation $x^\mu = x^\mu(\tilde{x}^\nu)$ are just only integral terms of total derivative, but it is changeless under local Lorentz transformation $e^{\hat{\alpha}}_\mu = \Lambda^{\hat{\alpha}}_{\hat{\beta}}(x) \tilde{e}^{\hat{\beta}}_\mu$.

## 2 Negative kinetic energy term in $L_{\mathrm{GK}}$ and its removing

We first discuss a simple case that all the three principal minors of the metric $g_{ij}$ are positive, e.g.,

$$g_{33} > 0, \quad \begin{vmatrix} g_{22} & g_{23} \\ g_{32} & g_{33} \end{vmatrix} > 0, \quad \left| g_{ij} \right| > 0. \tag{2-1}$$

An example of this case is the metric arisen from the ADM decomposition[2], which is indicated by the line element

$$\mathrm{d}s^2 = -\left( N^2 - h_{ij} N^i N^j \right) \left( \mathrm{d}x^0 \right)^2 + 2 N_i \mathrm{d}x^i \mathrm{d}x^0 + h_{ij} \mathrm{d}x^i \mathrm{d}x^j, \tag{2-2}$$

where $N_i = h_{ij} N^j$. For this case, $g_{ij} = h_{ij}$ and $h_{ij}$ satisfy (2-1).

For the case that $g_{ij}$ satisfy (2-1), we can introduce a group of new variables $w_u$, $u = 0,1,2,3,4,5$:

$$\mathrm{e}^{\sqrt{3} w_0} = \left| g_{ij} \right|, \quad \mathrm{e}^{\sqrt{3} w_1} = \frac{\sqrt{\left[ g_{22} g_{33} - (g_{23})^2 \right]^3}}{\left| g_{ij} \right|}, \quad \mathrm{e}^{w_2} = \frac{g_{33}}{\sqrt{g_{22} g_{33} - (g_{23})^2}},$$
$$w_3 = \frac{g_{23}}{g_{33}}, \quad w_4 = \frac{g_{23} g_{31} - g_{12} g_{33}}{g_{22} g_{33} - (g_{23})^2}, \quad w_5 = \frac{g_{12} g_{23} - g_{22} g_{31}}{g_{22} g_{33} - (g_{23})^2}; \tag{2-3}$$

from (2-3) we see that if there is not gravitation field, and $g_{11} = g_{22} = g_{33} = 1$, $g_{12} = g_{23} = g_{31} = 0$ (i.e., Minkowski metric), then $w_0 = w_1 = w_2 = w_3 = w_4 = w_5 = 0$.

Contrarily, according to (2-3) we can obtain easily the expression

$$g_{ij} = g_{ij}(w_0, w_1, w_2, w_3,\ w_4, w_5)\ . \tag{2-4}$$

Substituting (2-4) to (1-3), we obtain

$$\begin{aligned} L_{\mathrm{g}0} = \left(-g^{00}\right)\Big[ & -\frac{1}{2}\left(w_{0,0}\right)^2 + \frac{1}{2}\left(w_{1,0}\right)^2 + \frac{1}{2}\left(w_{2,0}\right)^2 \\ & + \frac{1}{2}\mathrm{e}^{2w_2}\left(w_{3,0}\right)^2 + \frac{1}{2}\mathrm{e}^{\sqrt{3}w_1 - w_2}\left(w_{4,0}\right)^2 + \frac{1}{2}\mathrm{e}^{\sqrt{3}w_1 + w_2}\left(w_3 w_{4,0} + w_{5,0}\right)^2 \Big]. \end{aligned} \tag{2-5}$$

Eq. (2-3) is a slight improvement of the transformation (3-6) in Ref. [1]. The profit from the form of (2-5) is that, after canonical quantization, there is not the problem of operator ordering under some gauge conditions in the corresponding quantum theory. However, (2-5) shows clearly that the quadratic term of time derivative in $L_{\mathrm{g}0}$ is non-positive definitive; this characteristic is incongruous with theory of field. The methods removing the negative kinetic energy term $-\frac{1}{2}\left(w_{0,0}\right)^2$ and some corresponding characteristics have been discussed in Ref. [1].

On the other hand, (2-5) only holds under (2-1); and, further, if we employ vierbein formalism and choose "time gauge" $e^0_{\hat{a}} = 0\ \ (a = 1,2,3)$ for tetrad $e^{\hat{\alpha}}_{\mu}$, then according to $g_{\mu\nu} = e^{\hat{\alpha}}_{\mu} e_{\hat{\alpha}\nu}$ we can obtain (2-1) naturally (see Ref. [1]). We therefore now discuss general case of $e^0_{\hat{a}} \neq 0\ \ (a = 1,2,3)$.

The last expression of (1-10) can be seemed as a quadratic form about $Y_{\hat{a}\hat{b}}$. According to some theorems and computational methods about reducing a quadratic form to the diagonal form[3, 4], (1-10) can be written to the form

$$L_{\mathrm{GK}} = -K_0^2 + K_1^2 + \left(\frac{e^0_{\hat{0}}}{g^{00}}\right)^2 K_2^2 + \left(\frac{e^0_{\hat{0}}}{g^{00}}\right)^2 K_3^2 + \frac{1}{-g^{00}} K_4^2 + \frac{1}{-g^{00}} K_5^2\ , \tag{2-6}$$

where

$$\begin{aligned} K_0^2 &= \frac{1}{4\left(e^0_{\hat{0}}\right)^2} \frac{\left(\sqrt{\phi^2+8}+\phi\right)\left(\sqrt{\phi^2+8}+\phi+2\right)^2}{\phi^2\sqrt{\phi^2+8}\left(3\sqrt{\phi^2+8}+\phi+8\right)} \left(\Delta^{\hat{a}\hat{b}} Y_{\hat{a}\hat{b}}\right)^2 \\ &= \frac{1}{-g^{00}} \phi_1 \Delta^{\hat{a}\hat{b}} \Delta^{\hat{c}\hat{d}} Y_{\hat{a}\hat{b}} Y_{\hat{c}\hat{d}}\ , \end{aligned} \tag{2-7}$$

$$\begin{aligned} & \phi \equiv \frac{-g^{00}}{\left(e^0_{\hat{0}}\right)^2},\quad \Delta^{\hat{a}\hat{b}} \equiv \eta^{\hat{a}\hat{b}} + \frac{2}{\left(e^0_{\hat{0}}\right)^2\left(\sqrt{\phi^2+8}+\phi+2\right)} e^{\hat{a}0} e^{\hat{b}0}, \\ & \phi_1 \equiv \frac{\left(\sqrt{\phi^2+8}+\phi\right)\left(\sqrt{\phi^2+8}+\phi+2\right)^2}{4\phi\sqrt{\phi^2+8}\left(3\sqrt{\phi^2+8}+\phi+8\right)}. \end{aligned} \tag{2-8}$$

$K_\nu$ $(\nu = 1, 2, 3, 4, 5)$ can be found in Appendix of this paper.

According to the corresponding conclusions about reducing a quadratic form to the diagonal form[3, 4], the form of (2-6) is unique.

Because there are six dynamical equations in which there are second time derivative terms in the ten Einstein equations $R^{\mu\nu} - \frac{1}{2} g^{\mu\nu} R = \frac{8\pi G}{c^4} T^{\mu\nu}$, there are six independent kinetic variables in the theory, and, the right expression of (2-6) shows the six corresponding kinetic energy terms.

From (2-6) we see that if $g^{00} > 0$, then there are three positive and three negative kinetic energy terms in $L_{\mathrm{GK}}$, respectively; contrarily, if $g^{00} < 0$, then there are five positive and one negative kinetic energy terms in $L_{\mathrm{GK}}$, respectively. For obtaining positive kinetic energy terms as much as possible, we choose

$$g^{00} = -\left(e^{\hat{0}0}\right)^2 + \left(e^{\hat{1}0}\right)^2 + \left(e^{\hat{2}0}\right)^2 + \left(e^{\hat{3}0}\right)^2 < 0 . \tag{2-9}$$

This is first condition for insuring positive definiteness of the kinetic energy term, which can be realized by choosing appropriate coordinate conditions or gauge conditions provided by local Lorentz transformation.

Of course, even if we choose $g^{00} < 0$, there is still a negative kinetic energy term $-K_0^2$ in $L_{\mathrm{GK}}$. For insuring positive definiteness of the kinetic energy term, we must choose some of ten gauge conditions for 16 variables $e_\mu^{\hat{\alpha}}$ such that $K_0 = 0$, namely,

$$\Delta^{\hat{a}\hat{b}} Y_{\hat{a}\hat{b}} = 0 . \tag{2-10}$$

Among the ten gauge conditions, the four are provided by coordinate transformation $x^\mu = x^\mu(\tilde{x}^\nu)$ and the six are provided by local Lorentz transformation $e_\mu^{\hat{\alpha}} = \Lambda_{\hat{\beta}}^{\hat{\alpha}}(x) \tilde{e}_\mu^{\hat{\beta}}$.

We call (2-10) *the positive kinetic energy condition.*

For instance, in Ref. [1], it has been proved that the following four gauge conditions

$$\begin{cases} e_{\hat{a}}^0 = 0 \quad (a = 1, 2, 3), \\ \left( \sqrt{|g_{lm}|} \dfrac{g^{0\lambda}}{g^{00}} \right)_{,\lambda} = 0 \end{cases} \tag{2-11}$$

make that (2-10) holds. In (2-11), the last is coordinate condition, others are provided by local Lorentz transformation.

However, (2-11) is improper for some metric form (We shall discuss the Robertson-Walker metric as an example). On the other hand, using the following formulas:

$$\tilde{g}^{ij} \Gamma_{ij}^0 = \frac{\sqrt{-g^{00}}}{\sqrt{-g}} \left( \sqrt{|g_{lm}|} \frac{g^{0\lambda}}{g^{00}} \right)_{,\lambda} ,$$

$$\eta^{\hat{a}\hat{b}}\tilde{e}^i_{\hat{a}}\tilde{e}^j_{\hat{b}} = g^{ij} - g^{0i}\frac{e^j_{\hat{0}}}{e^0_{\hat{0}}} - g^{0j}\frac{e^i_{\hat{0}}}{e^0_{\hat{0}}} + g^{00}\frac{e^i_{\hat{0}}}{e^0_{\hat{0}}}\frac{e^j_{\hat{0}}}{e^0_{\hat{0}}},$$

$$e^{\hat{a}0}e^{\hat{b}0}\tilde{e}^i_{\hat{a}}\tilde{e}^j_{\hat{b}} = g^{0i}g^{0j} - g^{00}g^{0i}\frac{e^j_{\hat{0}}}{e^0_{\hat{0}}} - g^{00}g^{0j}\frac{e^i_{\hat{0}}}{e^0_{\hat{0}}} + \left(g^{00}\right)^2\frac{e^i_{\hat{0}}}{e^0_{\hat{0}}}\frac{e^j_{\hat{0}}}{e^0_{\hat{0}}},$$

the positive kinetic energy condition (2-10) can be written to the following two forms:

$$\frac{\sqrt{\left(-g^{00}\right)^3}}{\sqrt{-g}}\left(\sqrt{|g_{lm}|}\frac{g^{0\lambda}}{g^{00}}\right)_{,\lambda} = \frac{\sqrt{\phi^2+8}-\phi+2}{\sqrt{\phi^2+8}+\phi+2}e^0_{\hat{a}}e^0_{\hat{b}}Y^{\hat{a}\hat{b}}, \tag{2-12}$$

$$\frac{\sqrt{-g^{00}}}{\sqrt{-g}}\left(\sqrt{|g_{lm}|}\frac{g^{0\lambda}}{g^{00}}\right)_{,\lambda} = -\frac{\sqrt{\phi^2+8}-\phi+2}{\sqrt{\phi^2+8}+\phi+2}\left(\frac{g^{0i}g^{0j}}{g^{00}} - 2g^{0i}\frac{e^j_{\hat{0}}}{e^0_{\hat{0}}} - \phi\, e^i_{\hat{0}}e^j_{\hat{0}}\right)\Gamma^0_{ij}. \tag{2-13}$$

We see that (2-11) is a special case of (2-12), and from (2-13) we see that, for removing the negative kinetic energy term in (2-6) under the general case, what we need to do is only to choose $e^\mu_{\hat{0}}$ such that (2-13) holds; and, generally speaking, sixteen variables $\left\{e^\mu_{\hat{\alpha}}\right\}$ that satisfy eleven equations (i.e., $g^{\mu\nu} = e^\mu_{\hat{\alpha}}e^{\hat{\alpha}\nu}$ and (2-13) ) are existent.

For example, we consider the Robertson-Walker metric indicated by the line element

$$\mathrm{d}s^2 = -\mathrm{d}t^2 + R^2(t)\left[\frac{\mathrm{d}r^2}{1-kr^2} + r^2\left(\mathrm{d}\theta^2 + \sin^2\theta\mathrm{d}\varphi^2\right)\right],$$

which can be obtained by some investigation of symmetry but not solving the Einstein field equation, for which (2-13) becomes

$$\left(e^1_{\hat{0}}\right)^2\frac{1}{r^2}\frac{1}{1-kr^2} + \left(e^2_{\hat{0}}\right)^2 + \left(e^3_{\hat{0}}\right)^2\sin^2\theta = \frac{3\left(e^0_{\hat{0}}\right)^2}{\left(R(t)r\right)^2}\frac{\sqrt{1+8\left(e^0_{\hat{0}}\right)^4} + 2\left(e^0_{\hat{0}}\right)^2 + 1}{\sqrt{1+8\left(e^0_{\hat{0}}\right)^4} + 2\left(e^0_{\hat{0}}\right)^2 - 1}. \tag{2-14}$$

We can choose $e^\mu_{\hat{0}}$ such that (2-14) holds. Especially, we can prove that neither of two groups of gauge conditions $e^{\hat{a}}_0 = 0$ and $e^0_{\hat{a}} = 0$ ($a$ = 1, 2, 3) is appropriate for (2-14).

Substituting (2-10) to the Einstein equations expressed by tetrad $\left\{e^{\hat{\alpha}}_\mu\right\}$ (the concrete forms of which are the formulas (1-23) ~ (1-25) in Ref. [1]), we obtain *a special form of the Einstein equations under the positive kinetic energy condition (2-10)* (whose concrete forms no longer be written down here). Because the positive kinetic energy condition (2-10) leads to that the second time derivative term of one of six kinetic variables $g_{ij}$ vanishes, *the special form of the Einstein equations under the positive kinetic energy condition (2-10)* are divided into five equations of constraint in which there is not any second time derivative term and five dynamical equations in which there are second time derivative terms, this characteristic is different from the normal

Einstein field equation $R^{\mu\nu}-\frac{1}{2}g^{\mu\nu}R=\frac{8\pi G}{c^4}T^{\mu\nu}$ in which there are four equations of constraint and six dynamical equations.

The positive kinetic energy condition (2-10) leads to a reduction of one in the six kinetic variables, this is allowable because there is a gauge transformation such that (2-10) holds for tetrad $\{e_{\mu}^{\hat{\alpha}}\}$. Another example of this case is that, in the theory of the Yang-Mills field, although $A_3^a$ is a real kinetic variable, we can still choose so called space-axial gauge $A_3^a=0$ [5]. The reason that space-axial gauge holds is that there exists a gauge transformation such that $A_3^a=0$ holds for arbitrary gauge field $A_{\mu}^a$.

It is important that we can prove that the Euler-Lagrange equations corresponding the action

$$S_{(2)}=\frac{c^3}{16\pi G}\int\sqrt{-g}\,\mathrm{d}^4x\left(L_{\mathrm{GK}}-L_{\mathrm{GU}}+K_0^2\right) \tag{2-15}$$

are just *the special form of the Einstein equations under the positive kinetic energy condition (2-10)*, and, it is obvious that there is not any negative kinetic energy term in (2-15).

By another way we can obtain the conclusion that the positive kinetic energy condition (2-10) leads to a reduction of one in the six kinetic variables. If we calculate the momenta $\pi^{ij}$ conjugate to $g_{ij}$ according to the Lagrangian $\sqrt{-g}\left(L_{\mathrm{GK}}-L_{\mathrm{GU}}+K_0^2\right)$ in (2-15) *formally*, [So called "formally" has two meanings: ① If we want to obtain the corresponding Hamiltonian representation from the action (2-15), then we must employ the ADM decomposition to realize 3+1 dimensional decomposition of space-time manifold at beginning. However, the metric arisen from the ADM decomposition has been included in the case that $g_{ij}$ satisfy (2-1), hence, the momenta conjugate to $g_{ij}$ what we calculate here are only formal; ② The action (2-15) is in fact about tetrad $e_{\mu}^{\hat{\alpha}}$, because $K_0^2$ is expressed by tetrad. Hence, if we want to calculate the momenta conjugate to $g_{ij}$ from the Lagrangian $\sqrt{-g}\left(L_{\mathrm{GK}}-L_{\mathrm{GU}}+K_0^2\right)$ in (2-15), then what we calculate are only formal.] then according to (1-7), (1-10), (1-11) and (2-7) we have

$$\pi^{ij}=\frac{\partial\left(\sqrt{-g}\left(L_{\mathrm{GK}}-L_{\mathrm{GU}}+K_0^2\right)\right)}{\partial g_{ij,0}}=\frac{\sqrt{-g}}{-g^{00}}\left(Q^{\hat{a}\hat{b}\hat{c}\hat{d}}+\phi_1\Delta^{\hat{a}\hat{b}}\Delta^{\hat{c}\hat{d}}\right)\left(\tilde{e}_{\hat{a}}^{l}\tilde{e}_{\hat{b}}^{m}\frac{\partial\Gamma_{lm}^{0}}{\partial g_{ij,0}}Y_{\hat{c}\hat{d}}+Y_{\hat{a}\hat{b}}\tilde{e}_{\hat{c}}^{l}\tilde{e}_{\hat{d}}^{m}\frac{\partial\Gamma_{lm}^{0}}{\partial g_{ij,0}}\right),$$

we therefore obtain

$$\pi^{\hat{a}\hat{b}}=e_i^{\hat{a}}e_j^{\hat{b}}\pi^{ij}=\sqrt{-g}\left(Q^{\hat{a}\hat{b}\hat{c}\hat{d}}+\phi_1\Delta^{\hat{a}\hat{b}}\Delta^{\hat{c}\hat{d}}\right)Y_{\hat{c}\hat{d}}\,. \tag{2-16}$$

If we range $Y_{\hat{a}\hat{b}}$ to a column matrix $\boldsymbol{Y}=\left[Y_{\hat{1}\hat{1}},\,Y_{\hat{2}\hat{2}},Y_{\hat{3}\hat{3}},Y_{\hat{2}\hat{3}},Y_{\hat{3}\hat{1}},Y_{\hat{1}\hat{2}}\right]^{\mathrm{T}}$, then we can prove that the rank and the determinant of the $6\times6$ matrix $\overline{\boldsymbol{Q}}=\left[\sqrt{-g}\left(Q^{\hat{a}\hat{b}\hat{c}\hat{d}}+\phi_1\Delta^{\hat{a}\hat{b}}\Delta^{\hat{c}\hat{d}}\right)\right]$ are 5 and 0, respectively. Hence, apart from the four obvious constraints $\pi^{0\lambda}=\frac{\partial\left(\sqrt{-g}\left(L_{\mathrm{GK}}-L_{\mathrm{GU}}+K_0^2\right)\right)}{\partial g_{0\lambda,0}}=0$, we have one and only one constraint about $\pi^{\hat{a}\hat{b}}$ obtained from (2-16). Concretely, by $\boldsymbol{C}=[C_1, C_2, C_3, C_4, C_5, C_6]$ we denote the eigenvector corresponding the unique zero eigenvalue of the matrix $\overline{\boldsymbol{Q}}$, namely, which satisfies $\boldsymbol{C}\cdot\overline{\boldsymbol{Q}}=\boldsymbol{0}=[0,0,0,0,0,0]$, the constraint about $\pi^{\hat{a}\hat{b}}$ thus reads $\boldsymbol{C}\cdot\boldsymbol{\pi}=0$, where $\boldsymbol{\pi}=\left[\pi^{\hat{1}\hat{1}},\ \pi^{\hat{2}\hat{2}},\pi^{\hat{3}\hat{3}},\ \pi^{\hat{2}\hat{3}},\ \pi^{\hat{3}\hat{1}},\ \pi^{\hat{1}\hat{2}}\right]^{\mathrm{T}}$.

We emphasize again that the characteristic "the quadratic term of time derivative in the action of general relativity is non-positive definitive" does not denote that there is inconsistency in the structure of the theory of general relativity, it only shows that if we regard general relativity as a theory of field (e.g. tetrad field), then this characteristic is incongruous with theory of field.

However, if negative kinetic energy term was inacceptable, then the action (2-15) should be regarded as the start of the theory. And, further, it is sure that results obtained by *the special form of the Einstein equations under the positive kinetic energy condition (2-10)* in which there are at least five equations of constraint and at most five dynamical equations are different from that obtained by the normal form of the normal Einstein field equations $R^{\mu\nu}-\frac{1}{2}g^{\mu\nu}R=\frac{8\pi G}{c^4}T^{\mu\nu}$. This question will be studied further.

## Appendix

The forms of the five functions $K_v\ (v=1,2,3,4,5)$ in (2-6) are as follows:

$$K_1^2=\frac{\left(\sqrt{\phi^2+8}-\phi\right)\left(3\sqrt{\phi^2+8}+\phi+8\right)}{8\left(e_{\hat{0}}^0\right)^2\phi^2\sqrt{\phi^2+8}}\left(\left(\frac{2}{\sqrt{\phi^2+8}+\phi+2}\eta^{\hat{a}\hat{b}}-\frac{e^{\hat{a}0}e^{\hat{b}0}}{e^{\hat{c}0}e_{\hat{c}}^0}\right)Y_{\hat{a}\hat{b}}\right)^2, \tag{A-1}$$

$$\begin{aligned}K_2=&\frac{\sqrt{2}}{\left[\left(e_{\hat{1}}^0\right)^2+\left(e_{\hat{2}}^0\right)^2+\left(e_{\hat{3}}^0\right)^2\right]\sqrt{\left[\left(e_{\hat{2}}^0\right)^2-\left(e_{\hat{3}}^0\right)^2\right]^2+\left(e_{\hat{1}}^0\right)^2\left[\left(e_{\hat{2}}^0\right)^2+\left(e_{\hat{3}}^0\right)^2\right]}}\\&\times\Big\{-2\left(e_{\hat{1}}^0\right)^2e_{\hat{2}}^0e_{\hat{3}}^0Y_{\hat{1}\hat{1}}+e_{\hat{2}}^0e_{\hat{3}}^0\left[\left(e_{\hat{1}}^0\right)^2-\left(e_{\hat{2}}^0\right)^2+\left(e_{\hat{3}}^0\right)^2\right]Y_{\hat{2}\hat{2}}+e_{\hat{2}}^0e_{\hat{3}}^0\left[\left(e_{\hat{1}}^0\right)^2+\left(e_{\hat{2}}^0\right)^2-\left(e_{\hat{3}}^0\right)^2\right]Y_{\hat{3}\hat{3}}\\&+e_{\hat{1}}^0e_{\hat{3}}^0\left[\left(e_{\hat{1}}^0\right)^2-3\left(e_{\hat{2}}^0\right)^2+\left(e_{\hat{3}}^0\right)^2\right]Y_{\hat{1}\hat{2}}+e_{\hat{1}}^0e_{\hat{2}}^0\left[\left(e_{\hat{1}}^0\right)^2+\left(e_{\hat{2}}^0\right)^2-3\left(e_{\hat{3}}^0\right)^2\right]Y_{\hat{3}\hat{1}}\\&+\left[\left(\left(e_{\hat{2}}^0\right)^2-\left(e_{\hat{3}}^0\right)^2\right)^2+\left(e_{\hat{1}}^0\right)^2\left(\left(e_{\hat{2}}^0\right)^2+\left(e_{\hat{3}}^0\right)^2\right)\right]Y_{\hat{2}\hat{3}}\Big\},\end{aligned} \tag{A-2}$$

$$\begin{aligned}K_3=&\frac{\sqrt{2}}{\sqrt{\left(e_{\hat{1}}^0\right)^2+\left(e_{\hat{2}}^0\right)^2+\left(e_{\hat{3}}^0\right)^2}\sqrt{\left[\left(e_{\hat{2}}^0\right)^2-\left(e_{\hat{3}}^0\right)^2\right]^2+\left(e_{\hat{1}}^0\right)^2\left[\left(e_{\hat{2}}^0\right)^2+\left(e_{\hat{3}}^0\right)^2\right]}}\\&\times\Big\{e_{\hat{1}}^0\left[\left(e_{\hat{2}}^0\right)^2-\left(e_{\hat{3}}^0\right)^2\right]Y_{\hat{1}\hat{1}}-e_{\hat{1}}^0\left(e_{\hat{2}}^0\right)^2Y_{\hat{2}\hat{2}}+e_{\hat{1}}^0\left(e_{\hat{3}}^0\right)^2Y_{\hat{3}\hat{3}}\\&-e_{\hat{2}}^0\left[\left(e_{\hat{1}}^0\right)^2-\left(e_{\hat{2}}^0\right)^2+\left(e_{\hat{3}}^0\right)^2\right]Y_{\hat{1}\hat{2}}+e_{\hat{3}}^0\left[\left(e_{\hat{1}}^0\right)^2+\left(e_{\hat{2}}^0\right)^2-\left(e_{\hat{3}}^0\right)^2\right]Y_{\hat{3}\hat{1}}\Big\},\end{aligned} \tag{A-3}$$

$$K_4 = \frac{1}{\sqrt{2}\sqrt{\left(e_{\hat{1}}^0\right)^2 + \left(e_{\hat{2}}^0\right)^2}\sqrt{\left(e_{\hat{1}}^0\right)^2 + \left(e_{\hat{3}}^0\right)^2}\left[\left(e_{\hat{1}}^0\right)^2 + \left(e_{\hat{2}}^0\right)^2 + \left(e_{\hat{3}}^0\right)^2\right]}$$

$$\times\left\{e_{\hat{2}}^0 e_{\hat{3}}^0\left[2\left(e_{\hat{1}}^0\right)^2 + \left(e_{\hat{2}}^0\right)^2 + \left(e_{\hat{3}}^0\right)^2\right]Y_{\hat{1}\hat{1}} - e_{\hat{2}}^0 e_{\hat{3}}^0\left[\left(e_{\hat{1}}^0\right)^2 + \left(e_{\hat{3}}^0\right)^2\right]Y_{\hat{2}\hat{2}} - e_{\hat{2}}^0 e_{\hat{3}}^0\left[\left(e_{\hat{1}}^0\right)^2 + \left(e_{\hat{2}}^0\right)^2\right]Y_{\hat{3}\hat{3}}\right.$$

$$\left. - 2e_{\hat{1}}^0 e_{\hat{3}}^0\left[\left(e_{\hat{1}}^0\right)^2 + \left(e_{\hat{3}}^0\right)^2\right]Y_{\hat{1}\hat{2}} - 2e_{\hat{1}}^0 e_{\hat{2}}^0\left[\left(e_{\hat{1}}^0\right)^2 + \left(e_{\hat{2}}^0\right)^2\right]Y_{\hat{3}\hat{1}} + 2\left[\left(e_{\hat{1}}^0\right)^2 + \left(e_{\hat{2}}^0\right)^2\right]\left[\left(e_{\hat{1}}^0\right)^2 + \left(e_{\hat{3}}^0\right)^2\right]Y_{\hat{2}\hat{3}}\right\},$$

(A-4)

$$K_5 = \frac{1}{\sqrt{2}\sqrt{\left(e_{\hat{1}}^0\right)^2 + \left(e_{\hat{2}}^0\right)^2}\sqrt{\left(e_{\hat{1}}^0\right)^2 + \left(e_{\hat{3}}^0\right)^2}\sqrt{\left(e_{\hat{1}}^0\right)^2 + \left(e_{\hat{2}}^0\right)^2 + \left(e_{\hat{3}}^0\right)^2}}$$

$$\times\left\{- e_{\hat{1}}^0\left[\left(e_{\hat{2}}^0\right)^2 - \left(e_{\hat{3}}^0\right)^2\right]Y_{\hat{1}\hat{1}} - e_{\hat{1}}^0\left[\left(e_{\hat{1}}^0\right)^2 + \left(e_{\hat{3}}^0\right)^2\right]Y_{\hat{2}\hat{2}} + e_{\hat{1}}^0\left[\left(e_{\hat{1}}^0\right)^2 + \left(e_{\hat{2}}^0\right)^2\right]Y_{\hat{3}\hat{3}}\right. \qquad \text{(A-5)}$$

$$\left. + 2e_{\hat{2}}^0\left[\left(e_{\hat{1}}^0\right)^2 + \left(e_{\hat{3}}^0\right)^2\right]Y_{\hat{1}\hat{2}} - 2e_{\hat{3}}^0\left[\left(e_{\hat{1}}^0\right)^2 + \left(e_{\hat{2}}^0\right)^2\right]Y_{\hat{3}\hat{1}}\right\}.$$

Under the "time gauge" $e_{\hat{a}}^0 = 0 \ \ (a = 1,2,3)$, all the functions $K_\nu \ (\nu = 1,2,3,4,5)$ are indeterminate forms $\left(\frac{0}{0}\right)$. For obtaining the values of $K_\nu \ (\nu = 1,2,3,4,5)$ under the limit case of $e_{\hat{a}}^0 = 0 \ \ (a = 1,2,3)$, we set

$$\begin{cases} e_{\hat{1}}^0 = \sqrt{e^{\hat{a}0} e_{\hat{a}}^0} \cos\theta, \\ e_{\hat{2}}^0 = \sqrt{e^{\hat{a}0} e_{\hat{a}}^0} \sin\theta\cos\varphi, \\ e_{\hat{3}}^0 = \sqrt{e^{\hat{a}0} e_{\hat{a}}^0} \sin\theta\sin\varphi; \end{cases}$$

Substituting the above expressions to (2-7), (A-1) ~ (A-5); and, further, taking $\sqrt{e^{\hat{a}0} e_{\hat{a}}^0} \to 0$, $\theta \to 0$ and $\varphi \to 0$, we obtain

$$K_0^2 = \frac{2}{3}\frac{1}{-g^{00}}\left(e^{\hat{a}i} e_{\hat{a}}^j \Gamma_{ij}^0\right)^2, \quad K_1^2 = \frac{1}{6}\frac{1}{-g^{00}}\left[\left(2e_{\hat{1}}^i e_{\hat{1}}^j - e_{\hat{2}}^i e_{\hat{2}}^j - e_{\hat{3}}^i e_{\hat{3}}^j\right)\Gamma_{ij}^0\right]^2,$$

$$K_2 = \sqrt{2} e_{\hat{1}}^i e_{\hat{3}}^j \Gamma_{ij}^0, \quad K_3 = -\sqrt{2} e_{\hat{1}}^i e_{\hat{2}}^j \Gamma_{ij}^0, \quad K_4 = \sqrt{2} e_{\hat{2}}^i e_{\hat{3}}^j \Gamma_{ij}^0, \quad K_5 = \frac{1}{\sqrt{2}}\left(- e_{\hat{2}}^i e_{\hat{2}}^j + e_{\hat{3}}^i e_{\hat{3}}^j\right)\Gamma_{ij}^0.$$

Substituting the above results to (2-6), we obtain the formula (3-3) in Ref. [1].